\documentclass[11pt]{article}
\usepackage{geometry}                % See geometry.pdf to learn the layout options. There are lots.
\usepackage{graphicx}
\usepackage{amssymb}
\usepackage{epstopdf}
\usepackage{authblk}
\usepackage{hyperref}
\hypersetup{
    colorlinks=true,
    linkcolor=cyan,
    filecolor=magenta,      
    urlcolor=blue,
}
\usepackage{comment,color}

\usepackage{ulem}

\begin{document}

\title{Modelling the spread of Covid19 in Italy using a revised version of the SIR model}
\author[1,2]{Andrea Palladino}
\author[1,3]{Vincenzo Nardelli}
\author[1]{Luigi Giuseppe Atzeni}
\author[1,5]{Nane Cantatore}
\author[1,6]{Maddalena Cataldo} 
\author[1,4]{Fabrizio Croccolo}
\author[1,7]{Nicolas Estrada}
\author[1,5]{Antonio Tombolini}
\affil[1]{CoVstat (Italy)}
\affil[2]{DESY, Platanenallee 6, Zeuthen (Germany)}
\affil[3]{Università degli Studi di Milano - Bicocca, Milano (Italy)}
\affil[4]{Universite de Pau et des Pays de l’Adour, E2S UPPA, CNRS, TOTAL, LFCR UMR5150, Anglet, (France)}
\affil[5]{Tombolini \& Associati srl, Loreto (Italy)}
\affil[6]{Università degli Studi dell'Aquila, L'Aquila, (Italy)}
\affil[7]{Università degli studi di Padova, Viccolo dell'Osservatorio 3, Padova (Italy)}

\date{May 2020}

\maketitle

\begin{abstract}
In this paper, we present a model to predict the spread of the Covid-19 epidemic and  apply it to the specific case of Italy. We started from a simple Susceptible, Infected, Recovered (SIR) model and we added the condition that, after a certain time, the basic reproduction number $R_0$ exponentially decays in time, as empirically suggested by world data. Using this model, we were able to reproduce the real behavior of the epidemic with an average error of 5\%. Moreover, we illustrate possible future scenarios, associated to different intervals of $R_0$.
This model has been used since the beginning of March 2020, predicting the Italian peak of the epidemic in April 2020 with about 100.000 detected active cases. The real peak of the epidemic happened on the 20th of April 2020, with 108.000 active cases. This result shows that the model had predictive power for the italian case.
\end{abstract}

\section{Introduction}
At the beginning of 2020, a previously-unknown respiratory tract disease was reported in China~\cite{Guo}. This event is having a huge negative impact worldwide, not only under an healthcare perspective, but also under the economic, social and cultural ones.
SARS-CoV-2 has been identified as the causative agent of the pandemic outbreak. 
It is a newly encountered member of the coronavirus family belonging to the RNA-viruses. Its behaviour is comparable to influenza viruses or SARS-CoV — the causative agent of the pandemic outbreak 2002/03 of \cite{Guo}\cite{Singhal}.
As soon as virus particles get into a host (human), they start invading cells (in this case, predominantly the ones in the respiratory tract) and these replicate the genome of the virus. Virus particles get into the host’s saliva and humans infect each other by talking to infected individuals, touching hands and close face-to-face interaction \cite{Singhal}\cite{Lai}.
A number that is a good landmark for the transmission rate, or the “infectiousness” of any infectious disease, is the basic reproduction number ($R_0$): this defines the average number of people that are infected by a single carrier over a defined period of time. 
$R_0$ is an indicator of the transmissibility of the epidemic and it has been defined as the average number of secondary cases that a single case can generate in a completely susceptible population. $R_0$ is of course dependent on the characteristics of the epidemic itself; however, it also depends on the population sample we are considering. The higher the human interaction in a population is, the higher the value $R_0$ will be. 
Among the factors that can influence $R_0$ in a given population there are therefore social habits and social organization. The basic reproduction number is  therefore mutable depending on these aspects, and the analysis of its variation in time can be crucial to monitor the trend in the transmissibility among a single population. For our specific case of study, i.e. the spread of the COVID-19 epidemic in Italy, it is important to study the variation of $R_0$ in time before and after the implementation of lockdown measures, as well as after its removal. 
In our modified SIR model, we let $R_0$ vary in time as a consequence of quarantine.  As we will show later, this will also allows us better reproduce real data, as well as have a predictive view on the whole trend of the epidemic.
% A second important parameter is the case fatality rate, or the "virus-associated mortality". A low fatality rate and many asymptomatic (but infectious) cases increase the basic reproduction number  \cite{Singhal} \cite{Wu} \cite{Li}.

Symptoms of coronavirus disease (COVID-19) are widespread: from asymptomatic patients to patients with flu-like symptoms up to a severe pneumonia leading to a severe acute respiratory distress symptom (ARDS) \cite{Guo}\cite{Singhal}, making the ventilation of patients unavoidable. 
%\added{Are there numbers how long ventilation is necessary? That would strengthen the point answer: from a medical point of view it is hard if not impossible to say, because of several reasons. In pubmed I cannot find publications treating the topic} 
%Considering the low fatality rate \cite{Guo} and its age adaptation, it is tempting to ease personal concerns about the new infectious disease but this can lead to a fatal outcome: Spreaders of the disease are young, healthy individuals with no or light symptoms \cite{Velavan}. But in a decompensating health system, not only the elderly and multimorbid individuals will lose out. 

Given the lack of reliable and long-term data regarding incubation period, virulence, contagiousness, and other transmission parameters \cite{Guo} for the novel coronavirus SARS-CoV-2 and the lack of reliable drugs and vaccines \cite{Lai}, containment measurements, the tracking of infected people and the treatment of patients in the early stage of the illness, remain the only feasible option to face the ongoing outbreak of the virus that is leading to a collapsing health system with thousands of deaths, as seen in hotspots.

The impact of Covid-19 in Italy has been and is still very severe, with a death toll notably high.
Up to now, more than 200.000 people have been tested positive in Italy and more than 30.000 died due to the Coronavirus. In this paper, we present the model used by the CoVstat group \cite{covstat} to model the spread of the epidemic. Using this model, we had been able to predict with good accuracy the peak of  the active infected, both related to its location in time (the date of the peak) and its amplitude (the maximum number of active infected).

\section{Methods} \label{sec:SIRmodels}

In this paragraph we present the SIR model, that is used as base for the development of further models from the CoVstat group. We point out that the SIR model is not predictive. In Sec.\ref{sec:beyondsir} we present the ingredients that represent a novelty here and have been included to improve the performance of the model and to describe the real data in a reasonable manner. 

\subsection{The standard SIR model}
SIR is one of the simplest models to describe the spread of an epidemic \cite{sir}. The SIR model is based on the assumption of a totally susceptible population at time $t_0$, i.e. the beginning of the spreading. In the SIR model, the overall population of \textbf{N} individuals is divided into 3 categories: susceptibles (\textbf{S}), infected (\textbf{I}) and removed (\textbf{R}). Hence, at a given time $t$ from the beginning of the spreading of the epidemic, $I(t)$ and $S(t)$ are the number of infected people present in the population and the number of vulnerable people that have not contracted the virus yet, respectively, while $R(t)$ is the sum of the ones that have developed immunity (recovered) or deceased and are therefore removed from the susceptible count. It is straightforward to notice that, at any time $t$, $S(t)+I(t)+R(t)=N$. The SIR model purpose is to describe the variation in time of S(t), I(t), and R(t) meaning the migration in time of individuals among these 3 categories.
The model consists of 3 categories: \textbf{S} for the susceptible people, \textbf{I} for the infected people and \textbf{R} for the sum of recovered and deceased people. 
The classic SIR model is described by 3 ordinary differential equations:
\begin{eqnarray*}
\frac{dS}{dt} &=& -\beta \frac{I \ S}{N} \\
\frac{dI}{dt} & = & \beta \frac{I \ S }{N} - \gamma I \\
\frac{dR}{dt} & = & \gamma I 
\end{eqnarray*}
%\sout{The condition $S+I+R = N$, where $N$ is constant and equal to the total population, is valid at any time.} 
where $\beta$ is related to the velocity of diffusion of the virus and $\gamma$ is related to the time required to infected people to become removed (recovered or deaths). Both parameters have the dimension of time$^{-1}$.
The three equations written above can be interpreted in the following manner:
\begin{itemize}
\item at the beginning of the epidemic, the entire population is susceptible to the infection ($S(0)=N$). If there is a single infected person, other people can get the infection, going from the category \textbf{S} to the category \textbf{I}. The strength (speed) of the spread of the virus is determined by the parameter $\beta$;
\item the number of infected people increases when susceptible people get infected. After a typical timescale equal to $1/\gamma$ infected people \textbf{I} go  in the third category \textbf{R}; 
\item the category \textbf{R} includes the sum of people that recovered or died after infection.
\end{itemize}
\begin{figure*}[t]
\centering
\includegraphics[width=0.48\textwidth]{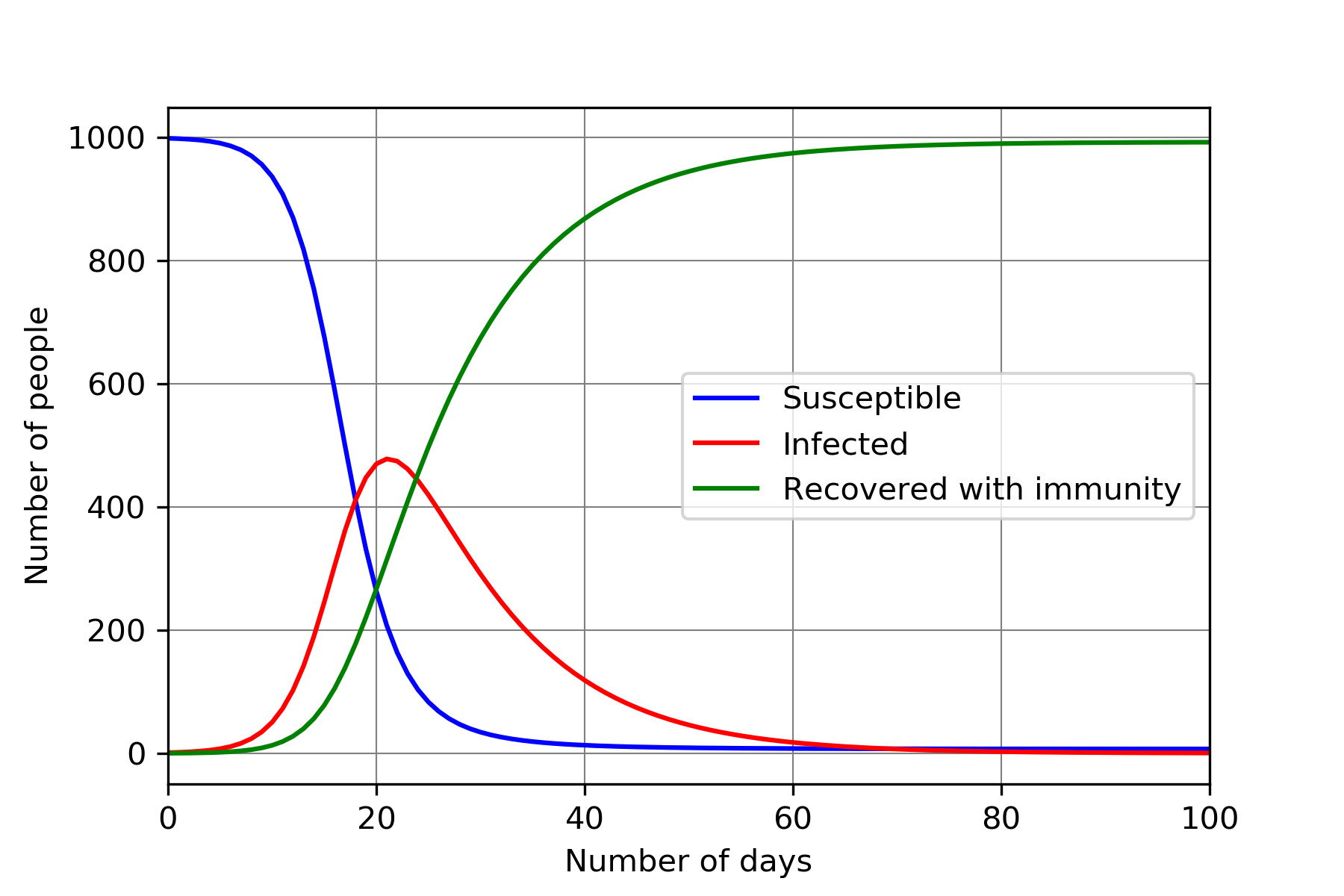}
\includegraphics[width=0.48\textwidth]{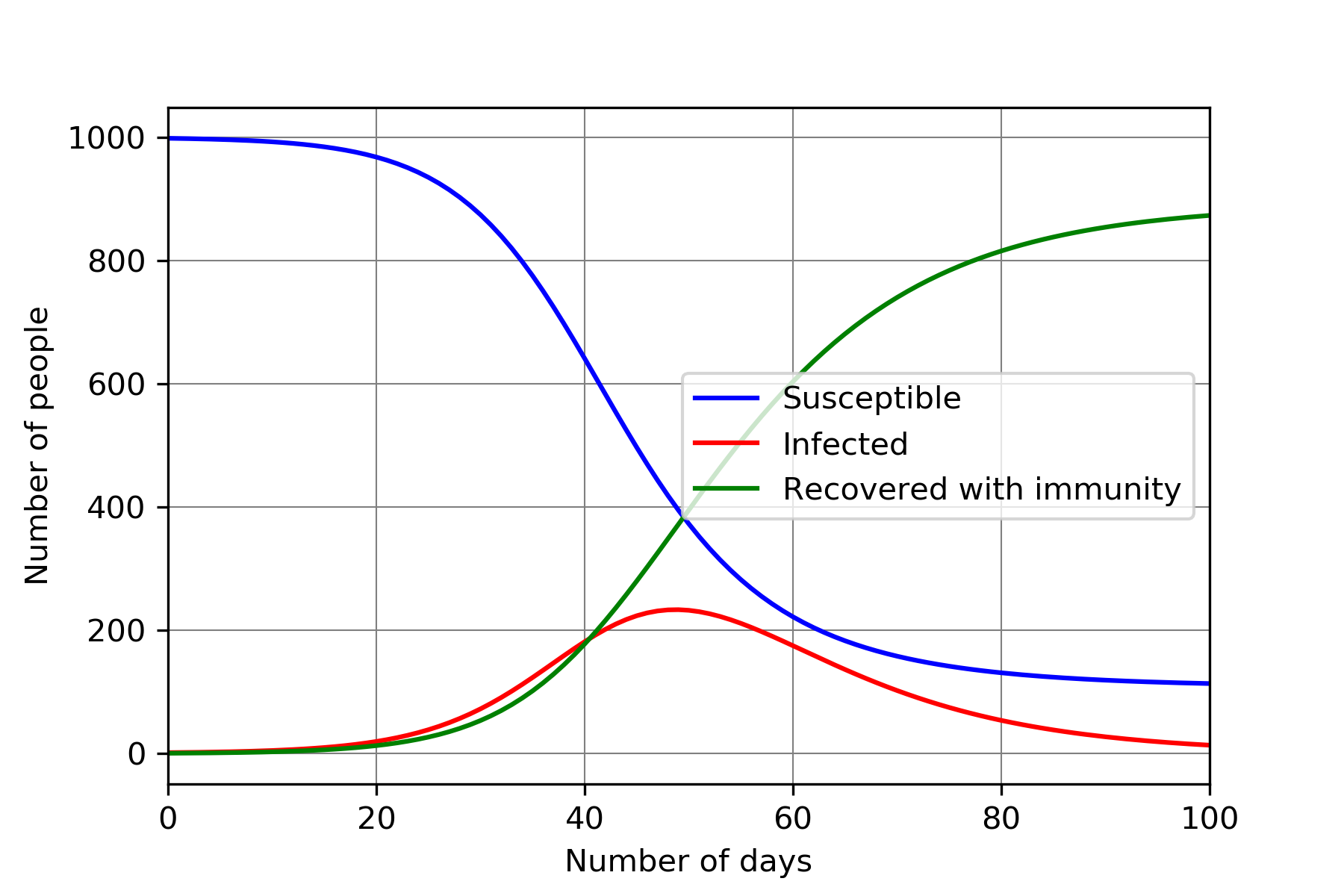}
\caption{Evolution of an epidemic using the SIR model, with $\gamma=\mathrm{0.1}$ and $\beta=\mathrm{0.5}$ on the left panel and $\beta=\mathrm{0.25}$ on the right panel. In the classical SIR model, $R_0$ is proportional to the fraction of susceptible people. Particularly it is equal to S/200 in the left panel and S/400 in the right panel. The condition $R_0=1$ occurs when the red curves reach their maximum.} 
\label{fig:sir}
\end{figure*}
Therefore, in the classic SIR model there are only 2 free parameters that can be used to fit real data: $\beta$ and $\gamma$. Both $\beta$ and $\gamma$ have dimension of time$^{-1}$. From here on, we will use days as unit of time.

In this model, the basic reproduction number $R_0$ is a dimensionless number obtained using a combination of the previous parameters:
$$
R_0(t) = \frac{\beta S(t)}{N \gamma}.
$$
$R_0(t)$ provides direct and quantitative information on the spread of the epidemic. If $R_0 \leq \mathrm{1}$ the epidemic will stop spontaneously, while with $R_0 > \mathrm{1} $ it will continue spreading. 
In Fig. \ref{fig:sir} we show two different evolution patterns of the pandemic, to demonstrate the effect of different values of $\beta$.
Both simulations start with the initial condition of 1 infected person, for a population of $N=$1000 people in total, and fixing $\gamma=\mathrm{0.1}$ (i.e.\ typical duration of the illness of 10 days). In both panels the blue line represents the number \textbf{S} of susceptible people which at time $t=0\ {\rm days}$ coincides with almost the entire population (for our simulation: \textbf{S}(0) = $1000-1$) and then decreases as both the recovered and infected numbers (green and red lines, respectively) increase. After an interval of time, in both the simulations, the infected number \textbf{I} reaches a peak, the value of which is dependent on the parameter $\beta$. After that time, the number of infected decreases as the recovered increase and ultimately reaches the total number of individuals.
In the simulation on the left we set $\beta=\mathrm{0.5}$ %where does that value actually come from? Is it based on real data? Then write that
while on the right we assume $\beta=\mathrm{0.25}$, to simulate the implementation of social distancing actions. We notice that in the left panel the peak of infected people comes after 20 days, with roughly half the population infected. On the right panel the peak is shifted and comes after 50 days and the number of infected people at the peak is less than the half of the previous case. This shows the importance of social distancing, %\textcolor{red}{ SE SI VEDE ANCHE R0 NEL GRAFICO SI PUO DISCUTERE ANCHE LA DIMINUZIONE DI R0 DOVUTA AL DISTANZIAMENTO }
since social distancing and quarantines reduce the parameter $\beta$, helping in reducing the number of infected people at the peak of the epidemic. This is indispensable to avoid the collapse of the healthcare system, and especially the intensive care units.
Although a simulation with the standard SIR appears to be adequate to describe an epidemic spreading in a sample where all the initial conditions remain constant throughout the period of time, it is not sufficient when it comes to a more complex and realistic situation such as the population of a given country, where the parameters of the model are influenced by other external factors. From this the necessity to modify the model for our case of study.
In the classic SIR model the parameter $\beta$ is constant, therefore it cannot account for the effect produced by quarantine actions, that would have also a dynamical impact on the parameter $R_0(t)$. 
% Moreover both recovered and deceased go in the same category and this does not permit to forecast the number of deaths at the end of the epidemic. 
For this reason, it is useful to go beyond the classic SIR model, as explained in the next section. 

\subsection{Beyond the classic SIR model}
\label{sec:beyondsir}
In the classic SIR model, the parameter $\beta$ is constant in time. This means that it cannot account for the slowdown of the spread due to the quarantine. To simulate a more realistic scenario, it is necessary to go beyond the classic SIR model. We denote this model as SIR 2.0. The open source code is available in the CoVstat repository \cite{repo}. Compared to the classic SIR it contains the following new features: 
\begin{itemize}
\item the parameter $\beta(t)$ changes in time, to account for the effects due to quarantine and social distancing. Particularly $\beta(t)=\beta_0$ before a time $t_{th}$, while it exponentially decays for $t>t_{th}$:
$$
\beta(t)= \beta_0 \ e^{-(t-t_{th})/\tau}  \ \ \ \mbox{for } t \geq t_{th}
,$$
where $t_{th}$ and $\tau$ are two additional parameters of the model. The time $t_{th}$ represents the starting time of the quarantine actions, while $\tau$ refers to the decaying period and it has the dimension of time. The previous assumption is driven by empirical observations, since this behavior of $R_0(t)$ has been observed in several countries during the Covid19 pandemic;

\item we take into consideration the possible presence of asymptomatic patients. The virologist Ilaria Capua has suggested that 2/3 of patients in Italy might be asymptomatic \cite{capua}. The study conducted on the population of Vo' Euganeo \cite{voeuganeo} reaches similar conclusions. Therefore, it is reasonable to assume that the total number of infected people is roughly 3 times higher that the number of people that were tested positive.
\end{itemize}
Compared to the classic SIR model, in this revised version there are 2 more parameters, $t_{th}$ and $\tau$. Therefore the SIR 2.0 is characterized by 4 parameters in total: $\beta_0$, $\gamma$,  $t_{th}$ and $\tau$. 

\section{Results}
In this paper we present and discuss varios results obtained using the model described above. We first focus on the curve of the active infected in Italy. Then we discuss possible future scenarios and how to compute the parameter $R_0$ for Italy and italian regions. 
From here on, we will always use data from the 24th of February to the 17th of May.

\subsection{The Italian case}
\label{sec:itacase}
\begin{figure*}[t]
    \centering
    \includegraphics[width=0.8\textwidth]{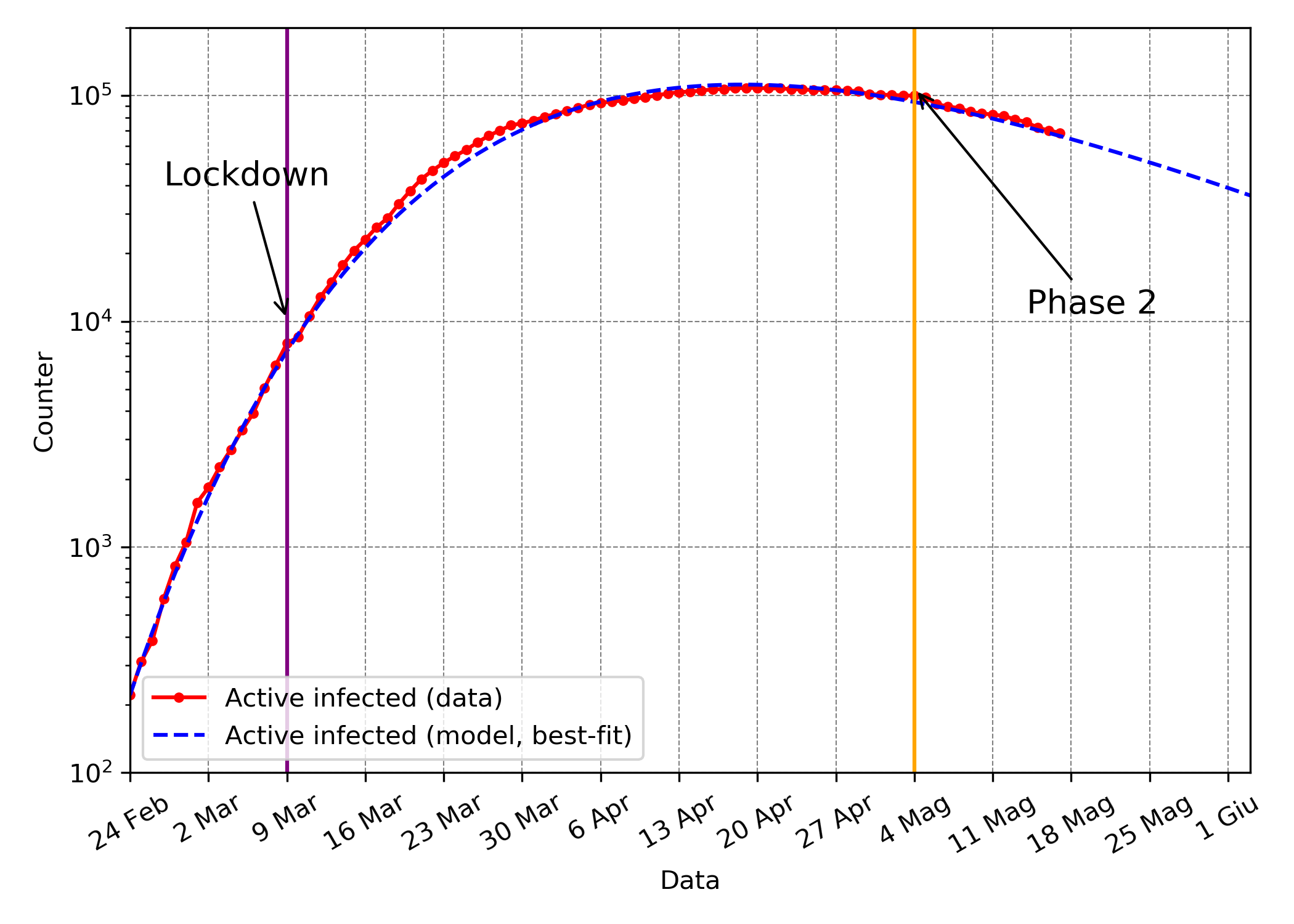}
    \caption{Simulation of the Italian case using the SIR 2.0 model. The red points represent the real number of active cases, while the blue curve represent the best-fit curve of the model. The magenta vertical line indicates the beginning of the quarantine, while the orange vertical line indicates the end of it.}
    \label{fig:SIR_ITA}
\end{figure*}

Using the model SIR 2.0 it is possible to reasonably describe the spread of COVID-19 in Italy. We used the data of Protezione Civile \cite{protciv}, starting on February 24th,  2020 (therefore, this  date corresponds to time zero of our model). We use $N=60.36 \times 10^6$ people as population and, as initial conditions, we choose $R_0=0$ and $I_0=221$, corresponding to the number of infected people on the 24th of February. 
In order to find the best model, we minimize the mean squared error between predictions and true data, allowing the 4 free parameters to vary. We consider as a last update the 17th of May.
The best fit is given by the following set of parameters:
$$
 \beta_0=0.384, \ \ \ \gamma=0.048, \ \ \ t_{th}=0, \ \ \  \tau=26.33\ .
$$
The best model has an average error of 4.9\% compared to the data and is shown in Fig.\ref{fig:SIR_ITA}  as the dashed blue line.
On the home page of \cite{covstat} the best fit model is updated daily. However it is important to recall that the predictions during the pandemic were stable. Since the beginning of March the model has predicted the peak of the infected people in April, with 100.000 of infected people (see Palladino's talk \cite{desytalk}). 

\subsection{Future predictions}
\begin{figure}[t]
\centering
\includegraphics[width=0.8\textwidth]{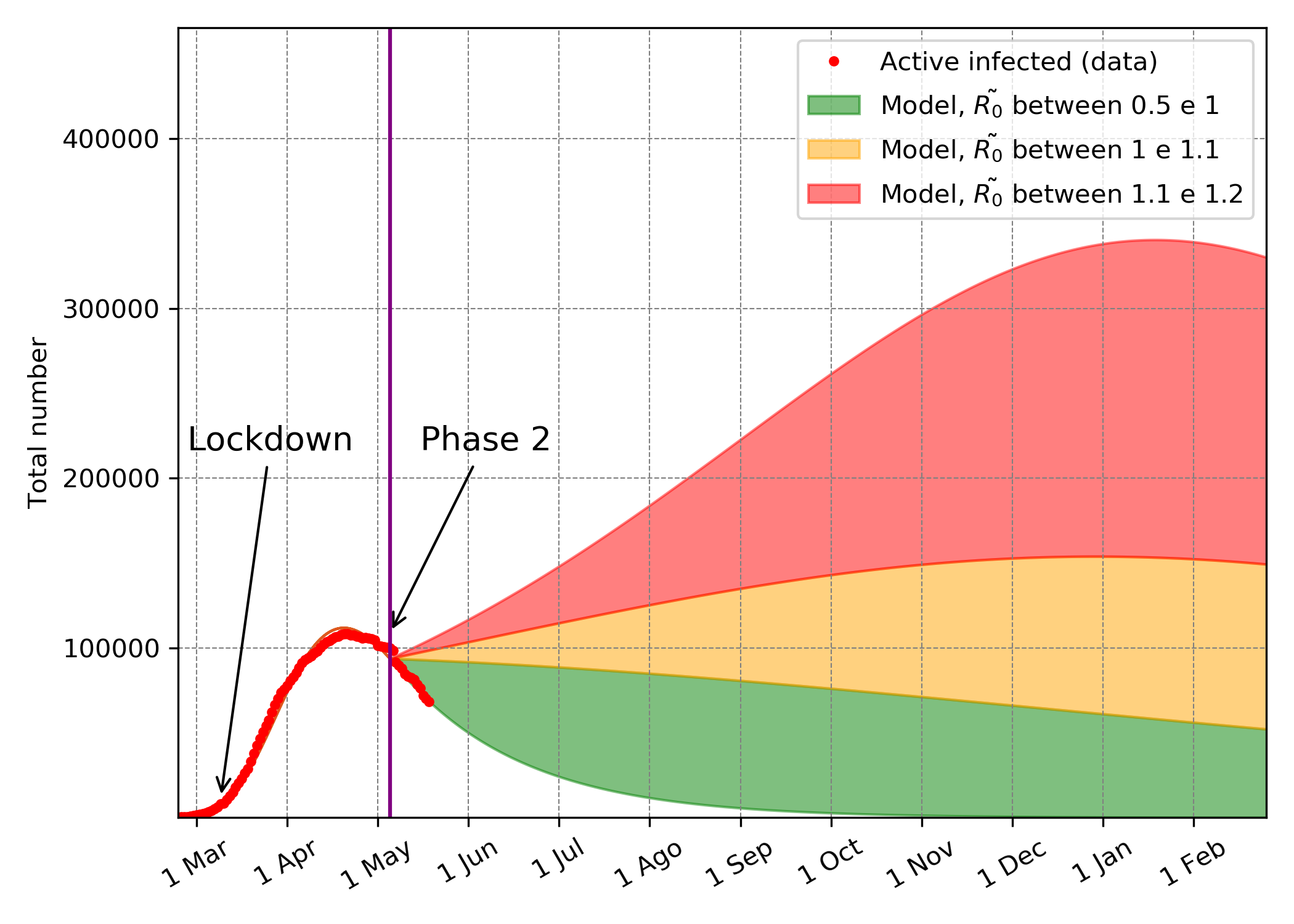}
\caption{Simulation of future scenarios, fixing a certain value of $R_0$ at the 4th of May. We identify 3 regions, corresponding to low risk (green), mid risk (orange) and high risk (red) situations.}
\label{fig:predictions}
\end{figure}
The phase 2 started in Italy on the 4th of May with progressive removal of quarantine. The conditions are different from phase 1, however, since the lockdown has not been completely released we cannot assume the initial conditions are fully restored. %\sout{Since the lockdown has been eased, the conditions are not the same anymore \textcolor{magenta}{(same idea as blue phrase before)}}. 
Even when the country's lockdown will formally be over, the local legislation about social distancing and the civic consciousness of population will be dramatically different than before. On the other side we still don't have enough information about the immunity for those who have contracted the virus. Therefore, it is hard, or even impossible, to make accurate predictions without a reasonable model for the evolution of $R_0(t)$. However, it is interesting to understand which future scenarios are associated to different intervals of the parameter $R_0(t)$. 

In order to do that, we fit the past data with the SIR 2.0, as explained in the previous section. Then, we use the basic SIR model for new future predictions, fixing a certain ratio $\beta(t)/\gamma$ on the last day of data. Therefore, the value of $\tilde{R_0}$ at the $t_d$=4th May is given by:
$$
\tilde{R_0} = \frac{\beta(t_d) S(t_d)}{\gamma N}
$$
where $S(t_d)$ is the number of susceptible people on $t_d$. 
Then $R_0(t)$ evolves as:
$$
R_0(t)=\tilde{R_0} \frac{S(t)}{N}
$$
Since in our model we assume that the total number of infected people is 3 times higher that the number of tested positives $I(t)$, the previous expressions becomes equal to:
$$
R_0(t)=\tilde{R_0} \left(1-\frac{3 I(t)}{N} \right)
$$
The epidemic starts decreasing when $R_0(t) < 1$. This condition is always satisfied when $\tilde{R_0} < 1$. In Fig.\ref{fig:predictions} we report 3 different regions corresponding to 3 different intervals of $\tilde{R_0}$, as explained in the legend of the figure.

\subsection{The computation of R0(t) for Italy and regions}
\begin{figure}[t!]
\centering
\includegraphics[width=0.48\textwidth]{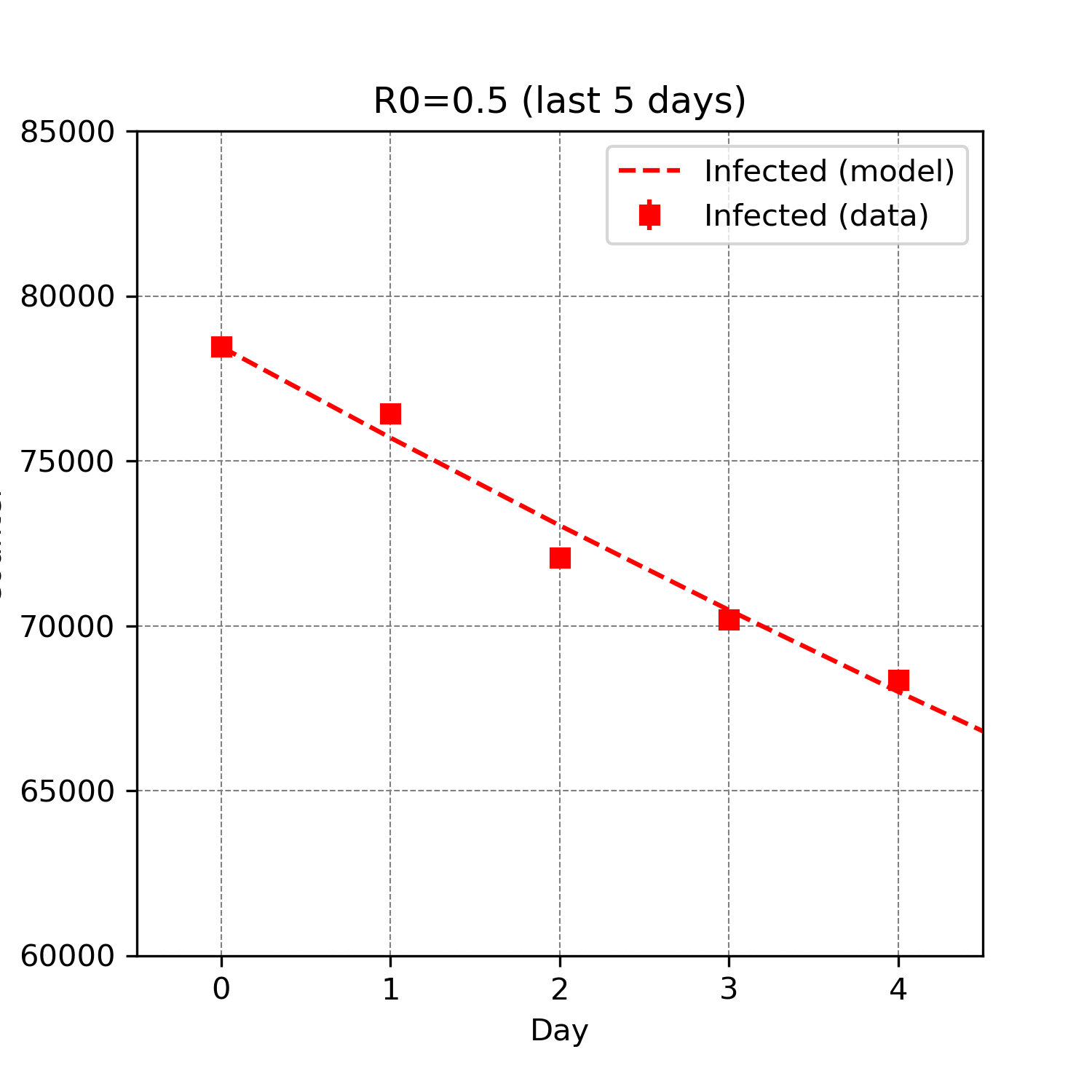}
\includegraphics[width=0.48\textwidth]{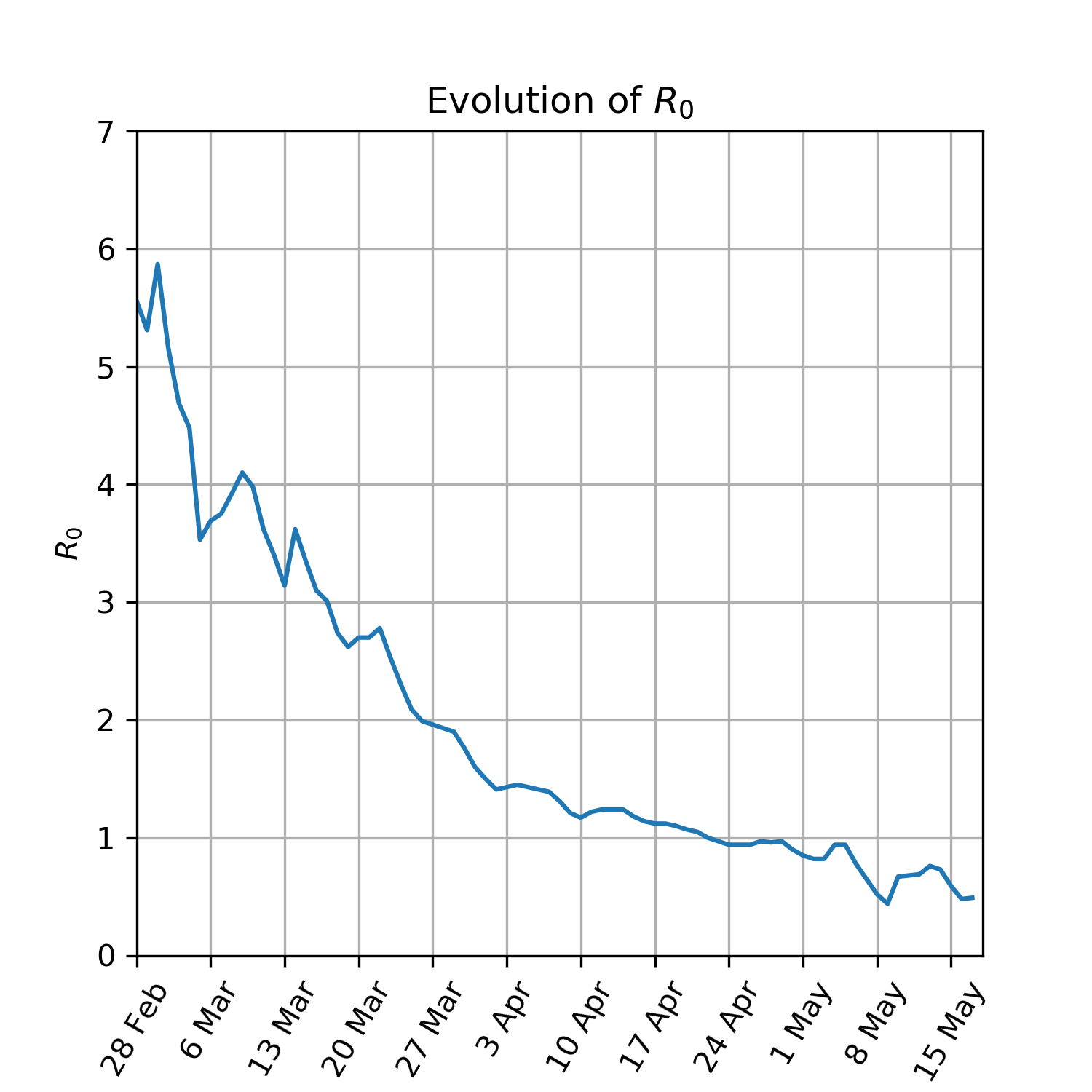} \\
\includegraphics[width=0.6\textwidth]{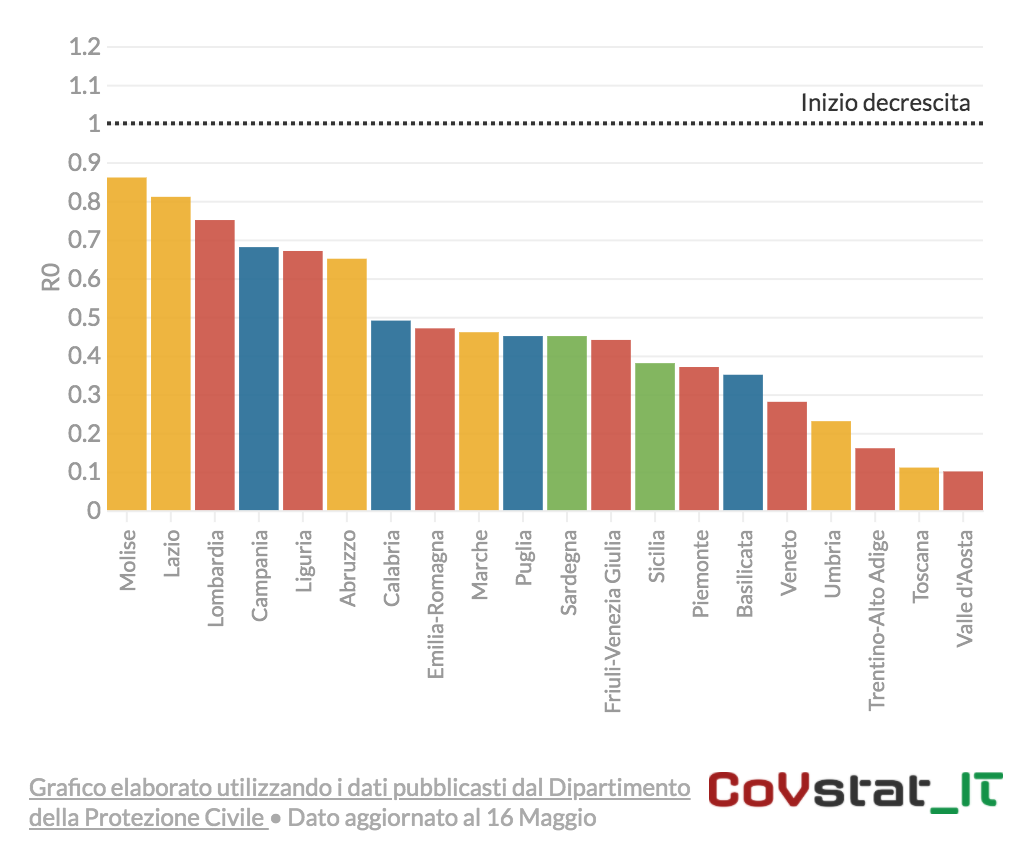}
\caption{Left panel: computation of $R_0(t)$ in Italy using the last 5 days of data. Right panel: evolution of $R_0(t)$ in Italy as function of time, using time window of 5 days. Mid panel: $R_0(t)$ computed for italian regions (updated to the 17th of May).}
\label{fig:r0}
\end{figure}

In this section we focus on the computation of $R_0(t)$ for Italy and Italian regions. In order to do that, we focus on a subset of data.  Particularly we use that last 5 days of data for Italy and last 7 days of data for Italian regions (due to the smaller statistics). This is done to avoid fast oscillations of $R_0$ and to better understand the general trend of the epidemic during the last week. In order to do that, we use the standard version of the SIR model. We assume an average duration of the illness of 14 days, i.e. $\gamma=1/14$. Then we minimize the mean squared errors between real data and model, varying $R_0(t)$, with the usual definition given for the SIR model.
Let us notice that, even assuming that the total number of infected people is 3 times larger than that the tested positive ones, the number of susceptible people is still very high.
Therefore the value of $R_0(t)$ corresponds, in very good approximation, to:
$$
R_0(t) \simeq \frac{\beta(t)}{\gamma}
$$
With the present numbers, this remains true even if the number of infected people were 10 times larger. Indeed up to now order of 200 K cases were tested positive. So, if the true number of infected people were 2.2 millions, the ratio $S(t_d)/N=0.97$, very close to 1. The evolution of $R_0(t)$ in Italy is shown on the right panel of Fig.\ref{fig:r0}. On the left panel of the same figure, we represent $R_0(t)$ computed on the last 5 days.

In the central panel of the same figure we report the last value of $R_0(t)$ for italian regions, updated to the 17th of May. In this case we use a time window of 7 days, to compensate the smaller statistics and to reduce the oscillations of $R_0(t)$. Using this procedure, $R_0(t)$ becomes a good indicator of the behavior of the pandemic during the last week.

\section{Conclusion}
We have built a model to predict the spread of the Covid-19 epidemic in Italy. We started from a simple SIR model and we added the condition that, after a threshold time, the basic reproduction number $R_0(t)$ exponentially decays in time, as empirically suggested by the spread of the epidemic in different countries. Using this model we were able to predict the peak of the epidemic 1.5 months before it happened, with an error of 1 week on the period and an error smaller than 10\% on the absolute numbers. We have also presented possible future scenarios, assuming different intervals of the parameter $R_0(t)$ after the 4th of May, i.e.\ when the lockdown in Italy has been released. We conclude explaining our procedure to compute $R_0(t)$, as a function of time, for Italy and Italian regions. This paper shows that the model has a good predictive power, when a period of quarantine is observed with fixed conditions. 

%\subsection{The role of asymptomatic patients}
%https://www.imperial.ac.uk/media/imperial-college/medicine/sph/ide/gida-fellowships/Imperial-College-COVID19-Europe-estimates-and-NPI-impact-30-03-2020.pdf

%\subsection{Future predictions}
%\newline 

\textbf{ACKNOWLEDGEMENTS} A.Palladino has received funding from the European Research Council (ERC) under the European Union’s Horizon 2020 research and innovation program (Grant No. 646623).

\end{document}